\documentstyle[pre,aps,epsf,multicol]{revtex}
\begin{document}

\title{On the critical behavior of the one dimensional diffusive pair 
contact process}
\author{G\'eza \'Odor}
\address{Research Institute for Technical Physics and Materials Science, \\
H-1525 Budapest, P.O.Box 49, Hungary}    
\maketitle

\begin{abstract}
The phase transition of the one-dimensional, diffusive pair contact process 
(PCPD) is investigated by $N$ cluster mean-field approximations and high 
precision simulations. The $N=3,4$ cluster approximations exhibit smooth 
transition line to absorbing state by varying the diffusion rate $D$ with 
$\beta_2=2$ mean-field order parameter exponent of the pair density.
This contradicts with former $N=2$ results, where two different mean-field
behavior was found along the transition line. 
Extensive dynamical simulations on $L=10^5$ lattices give estimates for
the order parameter exponents of the particles for $0.05 \le D \le 0.7$.
These data can support former two distinct class findings. However the gap 
between low and high $D$ exponents is narrower than estimated previously
and the possibility for interpreting numerical data as a single class 
behavior with exponents $\alpha=0.21(1)$, $\beta=0.41(1)$ assuming 
logarithmic corrections is shown. Finite size scaling and cluster simulation 
results are also presented.
\end{abstract}

\begin{multicols}{2}

\section{Introduction}
The exploration of nonequilibrium universality classes is current interest
of research. In this area most systems investigated exhibit phase transitions 
to absorbing states with such weak fluctuations from which no return is 
possible \cite{Dick-Mar,Hin2000}. For a long time only the robust directed
percolation (DP) universality class has been known \cite{Jan81,Gras82}. Later
systems with extra conservation laws and symmetries were shown to belong to
other universality classes \cite{Gras84,nekim,IJen94,woh}. 
In the past few years it turned out that there are novel classes in low 
dimensional reaction-diffusion systems where neither classical bosonic field 
theory nor symmetry arguments can give better understanding of the critical 
behavior \cite{dok}. This is probably 
due to the fact that in low dimensions topological constraints become 
effective, blocking the motion of reacting particles \cite{OdMe02}. While 
bosonic field theories can not capture this feature, fermionic field theories
have not been successful for such systems so far. In fact the critical behavior
of such models split according to fermionic or bosonic particles are involved
in \cite{Cardy-Tauber,barw2cikk,HT97,Carlon99}. 

Recently novel universal behavior is reported in some low-dimensional 
reaction-diffusion models featured by production at pairs and single
particle diffusion 
\cite{HT97,Carlon99,Hayepcpd,Odo00,HayeDP-ARW,coagcikk,binary,NP01,OSC02}.
In these systems the production compete with pair annihilation and diffusion. 
If production wins steady states with finite particle density appear in 
(fermionic) models with hard-core repulsion, while in unrestricted (bosonic) 
models the density diverges. By lowering the production/annihilation rate a 
doublet of absorbing states without symmetries emerges. 
One of such states is completely empty, the other possesses a single wandering 
particle. In case of fermionic systems the transition to absorbing states
is continuous with novel, yet not completely settled critical behavior.

The field theory \cite{HT97} describing bosonic particles could not be solved
by standard renormalization procedures, but hinted at a transition with
non-DP behavior. At the transition point of the 1d model it predicts a 
density decay of the form
\begin{equation}
\rho(t,p_c) \propto \left[ \frac{\ln(t)}{t} \right]^{1/2} \ ,
\end{equation}
while in the inactive phase: $\rho(t,p_c) \propto t^{-1/2}$. 
These were confirmed by simulations \cite{OdMe02}. 
In case of fermionic particles of this model
(PCPD) density matrix renormalization group analysis \cite{Carlon99}, coherent 
anomaly extrapolation \cite{Odo00} and simulations \cite{Hayepcpd,Odo00} 
found novel kind of critical phase transition. However the critical exponents 
seem to depend on the diffusion strength $D$ and different interpretations 
of data have been born.
These embrace the possibilities of continuously changing exponents, 
two-universality classes \cite{Odo00} and single class with huge corrections
\cite{Carlon99,MHUS}. 

Very recently well defined set of critical exponents are reported in different 
versions of binary production PCPD-like processes \cite{PK66}. However these
simulations were done at a fixed, high diffusion/annihilation rate and as will 
be shown in Sect. \ref{simu} the exponent estimates agree well with those 
of this paper in the high diffusion region. Even more recently two studies 
\cite{DM0207720,H0208345} reported non-universality in the dynamical behavior
of the PCPD. While the former one by Dickman and Menezes explored different
sectors (a reactive and a diffusive one) in the time evolution and gave 
nontrivial exponent estimates, the latter one by Hinrichsen provided a 
hypothesis that the ultimate long time behavior should be characterized by 
DP behavior. 

Just before the submission of this paper a preprint by Kockelkoren and
Chat\'e \cite{KC0208497} showed extensive simulation results for a modified 
version of PCPD that is in between fermionic and bosonic models. That means
that they discard the single particle occupation constraint on the lattice but
suppress multiple occupancy by an exponentially decreasing creation probability 
($p^{N/2}$) of the particle number. They claim that their stochastic 
cellular automaton (SCA) model shows smaller corrections to scaling than 
the PCPD and exhibit a single universality class transition. 

The two universality class scenario was backed by pair mean-field 
approximation \cite{Carlon99} that showed two different mean-field behavior 
by varying $D$ and simulations \cite{Odo00} for the order parameter density
exponents. Such kind of mean-field behavior is absent if we replace the
annihilation $AA\to\emptyset$ with coagulation $AA\to A$ \cite{coagcikk}.
By the investigation of the parity conserving version of the PCPD the 
mean-field and pair-mean-field approximations resulted in similar phase
diagram, but higher order cluster mean-field showed a single mean-field
class behavior \cite{OSC02} and the authors concluded that the for 
appropriate description of such binary production models at least $N=3$
clusters are needed. That mean-field behavior was indeed found in $d=d_c=2$
by simulations \cite{OSC02}.

In the present work I show $N=3,4$ cluster mean-field results for the
PCPD model that again suggest single mean-field universality class.
This does not necessarily imply that below $d_c=2$ only one class would
exist. Higher precision simulations than that of \cite{Odo00} are also
presented in the second part of this paper that provide better exponent
estimates but still leave this question open. A single universality 
class scenario may be accepted only if we assume logarithmic correction to
data.

\section{The PCPD model}

A PCPD like binary spreading process was introduced in an early work 
by Grassberger \cite{G82}. Its preliminary simulations in 1d showed 
a non-DP type transition, but these results have been forgotten for a 
long time. The diffusive pair contact process (PCPD) introduced by Carlon 
et al \cite{Carlon99} is controlled by two independent parameters: the 
probability of pair annihilation $p$ and the probability of particle 
diffusion $D$. The dynamical rules are
\begin{eqnarray}
AA\emptyset,\,\emptyset AA \rightarrow AAA  \qquad {\rm with \ rate}
\, & (1-p)(1-D)/2 \nonumber \\
AA \rightarrow \emptyset\emptyset \qquad  {\rm with \ rate}\,  &
p(1-D) \nonumber \\
A\emptyset \leftrightarrow \emptyset A \qquad {\rm with \ rate}\,  & D \ .
\label{DynamicRules}
\end{eqnarray}
The {\it site mean-field} approximation gives a continuous transition at 
$p=1/3$.
For $p \le p_c(D)$ the particle and pair densities exhibit singular behavior:
\begin{equation}
\rho(\infty,p)\propto (p_c-p)^{\beta} \qquad
\rho_2(\infty,p)\propto (p_c-p)^{\beta_2}
\end{equation}
while at $p = p_c(D)$ they decay as:
\begin{equation}
\rho(t,p_c) \propto t^{-\alpha} \ , \ \ \ 
\rho_2(t,p_c) \propto t^{-\alpha_2} \ \ \ ,
\end{equation}
with the exponents:
\begin{equation}
\alpha=1/2, \ \ \ \alpha_2=1, \ \ \ \beta=1, \ \ \ \beta_2=2 \ .
\end{equation}
\begin{figure}
\begin{center}
\epsfxsize=70mm
\centerline{\epsffile{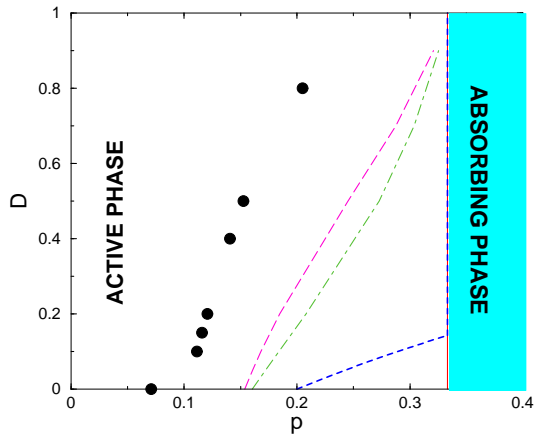}}
\caption{Schematic phase diagram of the 1d PCPD model.
Circles correspond to simulation and DMRG results, solid line
to site mean-field ($N=1$), dashed line to pair-approximation ($N=2$). 
Dot-dashed line shows $N=3$, long-dashed $N=4$ cluster mean-field 
results discussed in Sect.\ref{GMF}.}
\label{pcpd_pd}
\end{center}
\end{figure}
According to {\it pair mean-field} approximations the phase diagram can be
separated into two regions (see Fig.\ref{pcpd_pd}).
While for $D > 1/7$ the pair approximation gives the same $p_c(D)$ and 
exponents as the site mean-field, for $D\le 1/7$-s the transition line breaks 
and the exponents are different
\begin{equation}
\alpha=1, \ \ \ \alpha_2=1, \ \ \ \beta=1, \ \ \ \beta_2=1 \ .
\end{equation}
In the entire inactive phase the decay is characterized by the exponents:
\begin{equation}
\alpha=1, \ \ \ \alpha_2=2 \ .
\end{equation}

\section{Cluster mean-field results for PCPD} \label{GMF}

The generalized, {\it cluster mean-field} approximation introduced by 
\cite{gut87,dic88} was applied for the dynamical rules (\ref{DynamicRules})
of the 1d fermionic lattice model.
Master equations for $N=1,2,3,4$ block probabilities were set up
\begin{equation}
\frac{\partial P_N(\{s_i\})}{\partial t} = f\left (P_N(\{s_i\})\right) \ ,
\end{equation}
where site variables may take values: $s_i=\emptyset,A$. The equations could
be solved numerically for the $\frac{\partial P_N(\{s_i\})}{\partial t}=0$
steady state condition. Taking into account spatial reflection symmetries
of $P_N(\{s_i\})$ this involves 10 independent variables in case of $N=4$.
The particle ($\rho(p,D)$) and pair ($\rho_2(p,D)$) densities were expressed 
by $P_N(\{s_i\})$ and the phase transition point $p_c(D)$ was located for 
several values of $D$. At $p_c(D)$ quadratic fitting of the form
\begin{equation}
a ( p-p_c(D) ) + b ( p-p_c(D) )^2 \label{fit}
\end{equation}
was applied for $\rho(p,D)$ and $\rho_2(p,D)$.
The $N=1$ and $2$ solutions reproduced the results of \cite{Carlon99} 
for particle and pair densities. 
For $N=2$ the two regions, corresponding to different
leading order singularity of $\rho_2(p,D)$ with $\beta_2=1,2$ were located
by least square fit with the form (\ref{fit}).
For $N=3,4$ approximations smooth $p_c(d)$ phase transition lines are 
determined shown on Fig.\ref{pcpd_pd} and tabulated in Table \ref{GMFt}.
The quadratic fitting (\ref{fit}) resulted in leading order
singularities $\beta=1$ for particles and $\beta_2=2$ for pairs everywhere.
These are in contradiction with the $N=2$ approximation results similarly to
the parity conserving binary process model case \cite{OSC02}. For that model
simulations in 2d strengthened the single mean-field class behavior along
$p_c(D)$ and it was conjectured that the pair approximation is an odd one. 
Here again I conclude that at least $N>2$ level of approximation is 
necessary to obtain a correct mean-field behavior. 

The single mean-field class property does not necessarily mean that below 
$d_c$ a single class behavior should occur all along the $p_c(D)$ transition
line. For example in a similar model that exhibits an additional
global particle number conservation \cite{woh} such situation was found.
Therefore I investigated by extensive simulations this question.

\section{High precision simulation results} \label{simu}

The simulations were performed on $L=10^5$ sized rings with random 
sequential update version of PCPD evolving by the following rules. 
A particle and a direction are selected randomly. 
One of the following reaction is performed:
(a) a nearest neighbour exchange in the selected direction with 
probability $D$;
(b) an annihilation with the nearest neighbour particle in the selected 
direction with probability $p(1-D)$;
(c) a creation of a new particle in the selected direction at the
second nearest neighbour empty site with probability 
$(1-p)(1-D)$ if the nearest neighbour is filled with a particle.
\begin{figure}
\begin{center}
\epsfxsize=70mm
\centerline{\epsffile{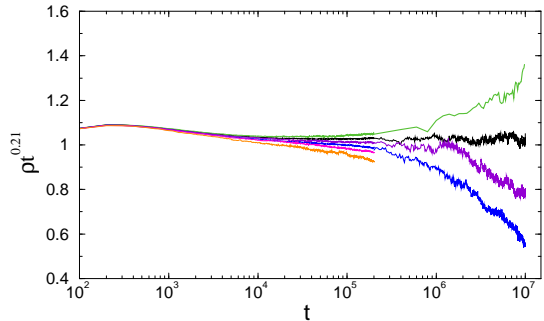}}
\caption{Density decay times $t^{0.21}$ in 1d PCPD at $D=0.7$ and 
$p=0.1574$,$0.15745$,$0.1575$,$0.15755$,$0.1576$,$0.1577$ 
(top to bottom curves). }
\label{dec7}
\end{center}
\end{figure}
The number of particles ($N_p$) is followed and the time is updated by 
$1/N_p$ following a reaction (throughout the whole paper the time is 
measured by Monte Carlo steps (MCS)).
The initial conditions were random distribution of particles with 
occupation probability $0.5$.

\begin{figure}
\begin{center}
\epsfxsize=70mm
\centerline{\epsffile{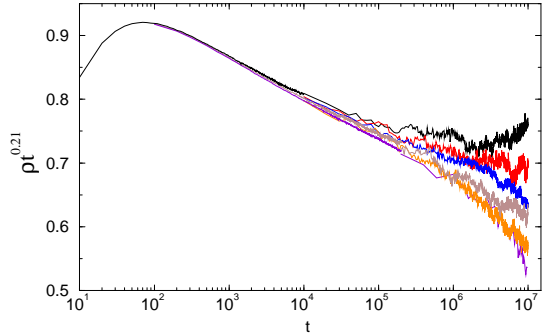}}
\caption{Density decay times $t^{0.21}$ in 1d PCPD at $D=0.1$ and 
$p=0.10686$,$0.10688$,$0.10689$,$0.1069$,$0.10691$,$0.10692$ (top to bottom).}
\label{dec1}
\end{center}
\end{figure}

\begin{figure}
\begin{center}
\epsfxsize=70mm
\centerline{\epsffile{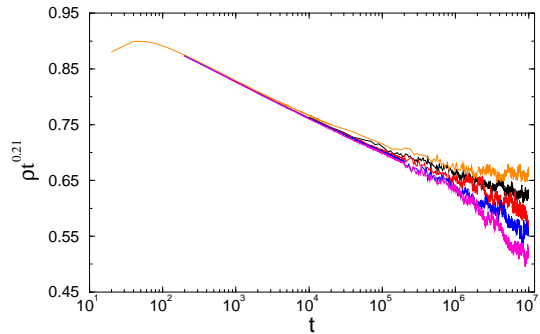}}
\caption{Density decay times $t^{0.21}$ in 1d PCPD for $D=0.05$ and
$p=0.10436$, $0.10438$,$0.1044$,$0.10441$,$0.10442$ (top to bottom). }
\label{dec05}
\end{center}
\end{figure}

It was suggested in \cite{DM0207720} that one may get smaller corrections to
scaling if one excludes the purely diffusive sector by averaging for states
having at least one pair in the system. In the present simulations I did not
find much effect (within statistical error margin) of such restrictions for 
the long time behavior.
\ref{dec1}, \ref{dec05}.
\begin{figure}
\begin{center}
\epsfxsize=70mm
\centerline{\epsffile{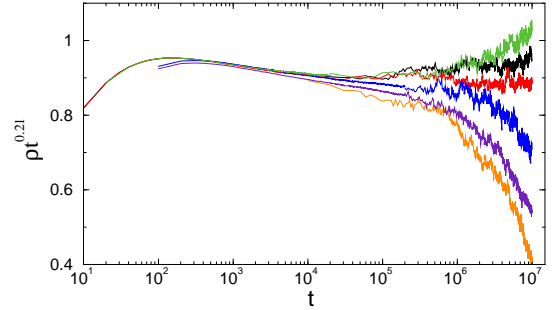}}
\caption{Density decay times $t^{0.21}$ in 1d PCPD at $D=0.5$ and 
$p=0.13351$, $0.13352$, $0.13353$, $0.13356$, $0.1336$, $0.13363$ 
(top to bottom). }
\label{dec5}
\end{center}
\end{figure}

\subsection{Density decay simulations} \label{denss}

\begin{figure}
\begin{center}
\epsfxsize=70mm
\centerline{\epsffile{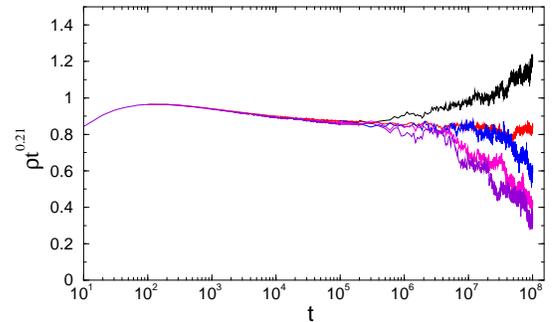}}
\caption{Density decay times $t^{0.21}$ in 1d PCPD at $D=0.2$ and 
$p=0.111215$,$0.11217$,$0.11218$,$0.11219$,$0.1122$ (top to bottom). }
\label{dec2}
\end{center}
\end{figure}

The critical point ($p_c$) for diffusion rates $D=0.05$, $0.1$, $0.2$, $0.5$, 
$0.7$ has been located by following the time evolution of the density decay. 
These simulations were done in two parts. First runs up to $t_{max}\sim 10^5$
MCS and with high statistical averages ($\sim 10^4$) were performed that 
allowed local slopes estimates of the density ($\rho(t)$) decay exponent 
$\alpha$ and $p_c$. 
These simulations were extended by long time runs up to $10^7-10^8$ MCS 
with $100-200$ sample numbers. The two sets of data are fitted 
together and are shown on Figs. \ref{dec7}, \ref{dec5}, \ref{dec2}, 
On all plots one can see up and down veering $\rho(t)$ curves in the long time
limit -- corresponding to active and absorbing phases -- separated by a 
roughly straight line -- corresponding to $p_c$. 
As one can see for high diffusion rates ($D\ge 0.2$) scaling with exponent
$\alpha \sim 0.21$ seems to a set for $t > \sim 3\times 10^4$ MCS.
This is in agreement with the first results provided for PCPD for high
diffusion rates \cite{Odo00} and with the results of 
\cite{HayeDP-ARW,DM0207720,KC0208497} for strong diffusions.
In cases $D=0.05$ and $0.1$ straight lines on the log-log plot appear from 
$t > \sim 3\times 10^2$ MCS with an exponent $\alpha=0.245(5)$.
This is in agreement with the results of \cite{PK66} who considered the case
when the coagulation and annihilation rate is three times the diffusion rate.
This exponent is about 10\% smaller than what was found in \cite{Odo00} 
but the two distinct class behavior seems to be supported.
Although the upper critical dimension of PCPD is expected to be at $d_c=2$
\cite{OSC02} one may not exclude the possibility of a second critical 
dimension ($d_c'=1$) or topological effects in 1d that may cause logarithmic 
corrections to scaling. For this reason I tried to apply logarithmic fitting
to the data of the form
\begin{equation}
\rho(t,p_c) = \left[ (a + b \ln(t)) / t \right]^{\alpha} \ \ .
\end{equation}
One can find the corresponding exponents in Table \ref{simtab} that are all
in agreement with the value $\alpha=0.21(1)$ in both the low and high diffusion
regions.
Here I applied least squares fitting for the most critical like curves 
such that the relative error in the sum of squares was at most 0.0001.
To confirm these results other critical exponents were investigated
using the precise $p_c$ values of this section.

\subsection{Steady state simulations} \label{sts}

To estimate directly the order parameter exponent describing the scaling 
\begin{equation}
\rho(\infty,\epsilon)\propto\epsilon^{\beta}
\end{equation}
off-critical, steady state densities had to be measured. Here again I used
$L=10^5$ system sizes.
The density decay was followed for each $D$ and $\epsilon_i = p_c-p_i$ values
on logarithmic time scales until saturation effect was observed.
Following that averaging of $\rho(t)$ was done for about $100$ samples 
within a time window that exceeds the saturation by a decade.
I measured the effective exponents like in \cite{Odo00} defined as
\begin{equation}
\beta_{eff} =  \frac { \ln( \rho(\infty,\epsilon_i) )
                    -\ln( \rho(\infty,\epsilon_{i-1})  ) }
                   { \ln( \epsilon_i)  - \ln( \epsilon_{i-1} ) }
\end{equation}
that are expected to converge to the true critical values in the
$\epsilon\to 0$ limit.
\begin{figure}[h]
\begin{center}
\epsfxsize=70mm
\centerline{\epsffile{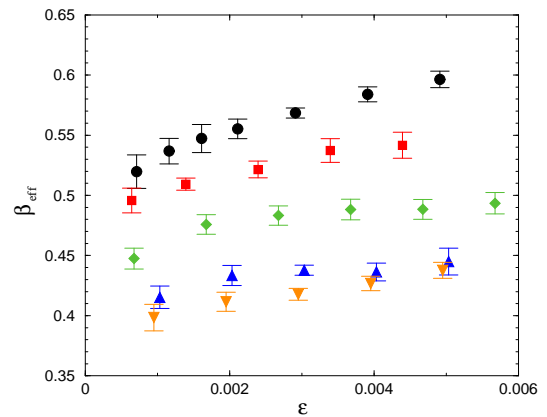}}
\caption{Effective $\beta$ exponents for different diffusion rates.
The circles correspond to $D=0.05$, the squares to $D=0.1$ the diamonds
to $D=0.2$, the up-triangles to $D=0.5$ and the down-triangles to $D=0.7$.}
\label{betaa}
\end{center}
\end{figure}
As one can see on Fig.\ref{betaa} the local slopes for $D=0.7$ and $D=0.5$
converge to $\beta=0.40(1)$ in agreement with the high diffusion rate results
provided in \cite{Odo00}. This value is also close to Hinrichsen's estimate
($0.38(6)$) for the cyclically coupled model \cite{HayeDP-ARW} and to 
Kockelkoren's value ($0.37(2)$) for the suppressed bosonic SCA model 
\cite{KC0208497}.

However for $D=0.05$ and $D=0.1$ extrapolations suggest $\beta=0.50(2)$. 
This is in agreement with Park's recent the results ($\sim 0.5$) \cite{PK66}
but somewhat off the low-diffusion data of ref. \cite{Odo00} 
($0.57(2)$) and from Dickman's estimates ($0.55-0.45$) \cite{DM0207720}.
The reason for these deviations is likely to be related to strong finite 
size effects, the complex way of scaling and the uncertainties of the $p_c$ 
values used.
\begin{figure}
\begin{center}
\epsfxsize=70mm
\centerline{\epsffile{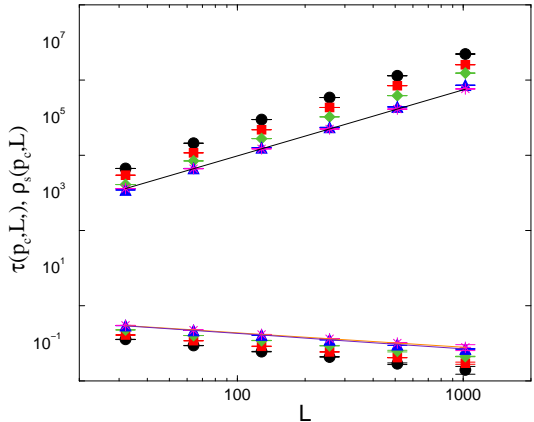}}
\caption{Finite size scaling of $\tau_L$ (upper points) and $\rho_L(\infty)$.
The circles correspond to $D=0.05$, the squares to $D=0.1$ the diamonds
to $D=0.2$, the up-triangles to $D=0.5$ and the down-triangles to $D=0.7$.
The lines show power-law fittings applied for $D=0.7$ data points.}
\label{fssZ}
\end{center}
\end{figure}
In case of the $D=0.2$ curve one may observe an extrapolation
to some intermediate value, but the curvature of the last points may also
suggest a tendency towards the high $D$ class data. 
Note that in earlier, lower scale simulations \cite{Odo00} the data for
$D=0.2$ showed low-$D$ critical behavior, strengthening the idea that 
some kind of very slow crossover happens here (although those results were
obtained for a SCA version of PCPD). 

Similarly to dynamical simulations I tried the possibility
if logarithmic correction to scaling
\begin{equation}
\rho(\infty,\epsilon) = \left[ \epsilon/ (a+ b\ln(\epsilon))\right]^{\beta} 
\label{logcor}
\end{equation}
could cure these ``uncertainties''. 
As one can see in Table \ref{simtab} the exponents for all $D$ values
satisfy scaling with $\beta=0.40(1)$ with logarithmic corrections of the 
form (\ref{logcor}).

\subsection{Finite size scaling}

Finite size scaling at the critical point was performed for system sizes
$L=32,64,128...1024$. The quasi-steady state density (averaged over surviving
samples) is expected to scale according to
\begin{equation}
\rho_s(\infty,p_c,L) \propto L^{-\beta/\nu_{\perp}} , \label{betapnfss}
\end{equation}
while the characteristic lifetime for half of the samples to reach the
absorbing state scales with the dynamical exponent $Z$ as
\begin{equation}
\tau(p_c,L) \propto L^Z \ .
\end{equation}
Since the system sizes are much smaller than in Sects. \ref{denss} and 
\ref{sts} one may expect stronger corrections to scaling. Indeed the power-law
fitting for $\beta/\nu_{\perp}$ results in values in the range $0.385-0.535$
and for $Z$ in the range $1.75-2$ depending on $D$. These results are shown
of Fig.\ref{fssZ}. Again the low-$D$ data are in agreement with those
of \cite{Carlon99}, \cite{PK66} and \cite{DM0207720}, while the high-$D$ data 
with those of \cite{Carlon99}, \cite{KC0208497} and \cite{HayeDP-ARW}. 
Just considering these ranges one can not distinguish this transition from
the PC class (with $\beta/\nu_{\perp}=0.500(5)$ and $Z=1.75(1)$ \cite{IJen94})
that caused initial debates in the literature \cite{Carlon99,Hayepcpd,Odo00}. 
Assuming single universality class corresponding to high-$D$ data 
we may expect: $\beta/\nu_{\perp}=0.38(1)$ and $Z=1.75(15)$.

\section{Conclusions}

In this paper I addressed the long standing question of diffusion dependence
of the phase transition of the PCPD model.
The $N=3,4$ level cluster mean-field calculations confirmed a single
mean-field universality class scenario similarly to the parity conserving
version of this model \cite{OSC02}. Again the best conclusion one can draw
from these data is that the $N=2$ pair approximation is an odd one and we need
at least $N>2$ level of mean-filed to get the correct scaling behavior
for binary production models.

The extensive simulations have confirmed at least one set of the exponents
-- those for high diffusion-- of the early results given in \cite{Odo00}.
Data in the low diffusion range are in good agreement with other recent
simulation results suggesting a different universality class. Although the 
scaling seems to set in much earlier in the low diffusion region than in
the high diffusion range, a slow crossover to high-$D$ behavior can be 
verified numerically assuming logarithmic corrections.
Similar conclusions can also be drawn from steady state simulation results.
Although the two universality class picture proposed in \cite{Odo00}
can not be excluded, data with logarithmic corrections may support a 
single class transition. Field theoretical arguments supporting or
excluding logarithmic corrections would be necessary. Note that in 1d
coupled systems logarithmic corrections are not rare at all.

The finite size simulations could not give decisive support for any of the
the possible dependence on diffusion of this transition, but the range of 
results are in agreement with those of other numerical results of the
literature. 

Mean-filed exponents, the upper critical dimension and the lack of time 
reversal symmetry in this model seem to exclude the possibility of further 
crossovers to an ultimate DP critical behavior. 
Finally the insensitivity to parity
conservation in binary production models brings up the question of 
insensitivity for other conservation laws, hence binary production,
diffusive models with global conservation might belong to the same class.

\vskip 0.5cm

\noindent
{\bf Acknowledgements:}\\
I thank H. Chat\'e and M. Henkel for useful discussions. 
Support from Hungarian research funds OTKA (Grant No. T-25286), Bolyai
(Grant No. BO/00142/99) and IKTA (Project No. 00111/2000) is acknowledged.
The simulations were performed on the parallel cluster of SZTAKI and on the
supercomputer of NIIF Hungary within the framework of the DEMOGRID project.

\begin{table}[h]
\begin{center}
\begin{tabular}{|c|c|c|c|c|c|c|c|c|c|}
$D$ & \multicolumn{3}{c|}{$N=2$} & \multicolumn{3}{c|}{$N=3$} & 
\multicolumn{3}{c|}{$N=4$}\\
    &  $p_c$  & $\beta $ & $\beta_2$ & $p_c$ & $\beta $ & $\beta_2$ & $p_c$ & 
$\beta $ & $\beta_2$ \\
\hline
0.9       &  0.3333& 1  & 2 &  0.3252 & 1  & 2 & 0.3208 & 1 & 2 \\   
0.7       &  0.3333& 1  & 2 &  0.3036 & 1  & 2 & 0.2875 & 1 & 2 \\
0.5       &  0.3333& 1  & 2 &  0.2727 & 1  & 2 & 0.2452 & 1 & 2 \\
0.2       &  0.3333& 1  & 2 &  0.2079 & 1  & 2 & 0.1845 & 1 & 2 \\
0.1       &  0.2888& 1  & 1 &  0.1840 & 1  & 2 & 0.1680 & 1 & 2 \\
0.05      &  0.2421& 1  & 1 &  0.1721 & 1  & 2 & 0.1606 & 1 & 2 \\
0.0002    &  0.2002& 1  & 1 &  0.1604 & 1  & 2 & 0.1537 & 1 & 2 \\
\end{tabular}
\end{center}
\caption{Summary of $N=2,3,4$ approximation results\label{GMFt}}
\end{table}

\begin{table}[h]
\begin{center}
\begin{tabular}{|c|c|c|c|}
$D$  &  $p_c$      & $\beta $ & $\alpha$ \\
\hline
0.7  &  0.15745(1) & 0.39(1)  & 0.214(5) \\
0.5  &  0.13353(1) & 0.414(16) & 0.206(7) \\
0.2  &  0.11218(1) & 0.402(8) & 0.217(8) \\
0.1  &  0.10688(1) & 0.407(7) & 0.206(7) \\
0.05 &  0.10439(1) & 0.411(10)& 0.216(9) \\
\end{tabular}
\end{center}
\caption{Summary of simulation results assuming logarithmic 
corrections\label{simtab}}
\end{table}

\end{multicols}
\end{document}